\documentclass[aps,showpacs,amsmath,amssymb,floatfix,onecolumn,superscriptaddress]{revtex4}

\usepackage{graphicx}
\usepackage{dcolumn}
\usepackage{bm}
\begin{document}
\newcommand{\Xnot}{X\textsuperscript{0}}
\newcommand{\Xm}{X\textsuperscript{-1}}
\newcommand{\Xmm}{X\textsuperscript{-2}}
\preprint{For submission to PRL}
\title{Optically Probing Spin and Charge Interactions in an Tunable Artificial Molecule}
\author{H. J. Krenner}
\affiliation{Walter Schottky Institut and Physik Department, Technische Universit\"at M\"unchen, Am Coulombwall 3, D-85748 Garching, Germany} 
\author{E. C. Clark}
\affiliation{Walter Schottky Institut and Physik Department, Technische Universit\"at M\"unchen, Am Coulombwall 3, D-85748 Garching, Germany} 
\author{T. Nakaoka}
\affiliation{Walter Schottky Institut and Physik Department, Technische Universit\"at M\"unchen, Am Coulombwall 3, D-85748 Garching, Germany} 
\author{M. Bichler}
\affiliation{Walter Schottky Institut and Physik Department, Technische Universit\"at M\"unchen, Am Coulombwall 3, D-85748 Garching, Germany} 
\author{C. Scheurer}
\affiliation{Lehrstuhl f\"ur Theoretische Chemie, Technische Universit\"at M\"unchen, Lichtenbergstrasse 4, D-85748 Garching, Germany} 
\author{G. Abstreiter}
\affiliation{Walter Schottky Institut and Physik Department, Technische Universit\"at M\"unchen, Am Coulombwall 3, D-85748 Garching, Germany} 
\author{J. J. Finley}
\affiliation{Walter Schottky Institut and Physik Department, Technische Universit\"at M\"unchen, Am Coulombwall 3, D-85748 Garching, Germany} 

\date{\today}
\begin{abstract}
We optically probe and electrically control a single artificial molecule containing a well defined number of electrons. Charge and spin dependent inter-dot quantum couplings are probed optically by adding a single electron-hole pair and detecting the emission from negatively charged exciton states. Coulomb and Pauli blockade effects are directly observed and tunnel coupling and electrostatic charging energies are independently measured. The inter-dot quantum coupling is shown to be mediated by electron tunneling.  Our results are in excellent accord with calculations that provide a complete picture of negative excitons and few electron states in quantum dot molecules.
\end{abstract}

\maketitle
Quantum bits based on charge and spin degrees of freedom in quantum dots (QDs) have attracted much attention over recent years since they can be electro-optically manipulated and read out\cite{Petta,tarucha,zrenner,stufler} and the extension to few dot systems with coupled quantum states is possible.\cite{krennerPRL,krennerNJP,ortner} Ultrafast optical gating of excitons in QDs has already been demonstrated by a number of groups.\cite{zrenner,stufler} However, radiative lifetimes are probably too short ($\sim 1$ ns) for excitons to be considered as viable quantum memory.  In comparison, the electron spin couples weakly to environmental degrees of freedom but requires local time dependent magnetic or electric fields for manipulation.\cite{Petta,Kroutvar04,Elzerman04} These properties have led to a number of mixed strategies whereby quantum memory is based on the electron spin and qubit-qubit coupling and ultrafast optical gating is mediated by via charged exciton auxiliary states.\cite{Troiani03,Calarco03}  However, despite theoretical progress\cite{Troiani03,Calarco03,szafran} no observations of negatively charged excitons in few-QD systems have been reported until now. 
\\  
In this letter we optically probe and electrically tune quantum couplings between charges and spins in a two atom QD-molecule (QDM).  As the number of resident electrons is varied from $n_e=0,1$ to $2$ each $n_e$-electron state is tuned into resonance at a distinct electric field, whereupon, the electrons hybridize and delocalize over the two dots forming molecular like quantum states.  These effects are probed by adding a single exciton (\Xnot=1$e$+1$h$) and detecting emission from negatively charged excitons.  Coulomb and Pauli blockade phenomena are directly observed in both exciton initial and few-electron final states.  Inter-dot tunnel couplings and charging energies are measured independently for different charge states and our findings are in very good agreement with calculations.
\\
\begin{figure}
	\centering
		\includegraphics[width=.700\columnwidth]{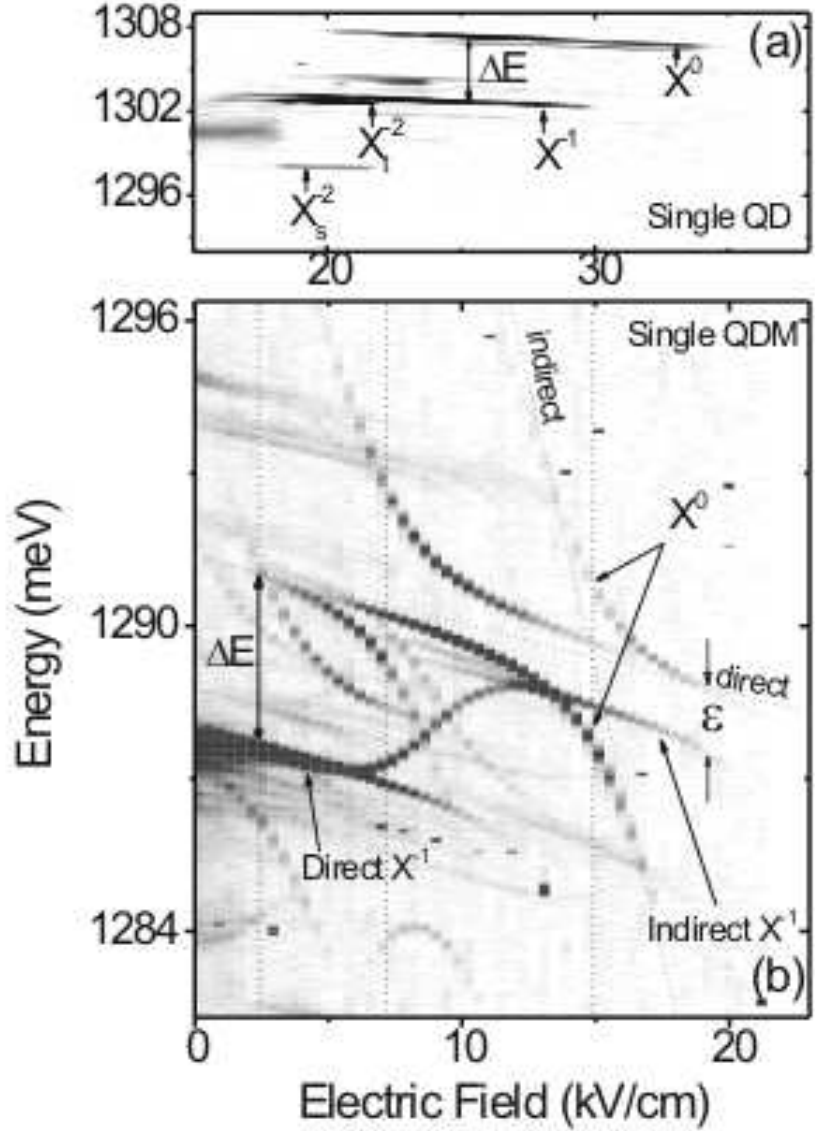}
	\caption{Comparison of electric field dependent PL recorded from a single InGaAs QD as a function of charge status (a) and an individual QDM (b).  Recombination from the charge neutral single exciton (\Xnot) is prominent at high electric fields and is progressively replaced by negatively charged excitons (\Xm, \Xmm) as the electric field reduces.}
	\label{fig1}
\end{figure}
The devices investigated consisted of a $\mathrm{In_{0.5}Ga_{0.5}As}$ self-assembled QD-molecule (QDM) embedded within a GaAs $n-i$ Schottky photodiode.  By applying a voltage between the $n$-doped contact and the Schottky-gate, the static electric field $(F)$ can be tuned from zero up to $\sim$35 kV/cm.\cite{krennerNJP} The two dots forming the molecule were nominally separated by a $d = 10$ nm thick GaAs spacer layer defining the intrinsic strength of the inter-dot tunnel coupling \cite{krennerNJP,krennerPRL}.
\\
The greyscale plot presented in fig \ref{fig1} compares PL recorded as a function of electric field from an uncoupled $\mathrm{In_{0.5}Ga_{0.5}As}$ QD \cite{baier} in a similar device (fig \ref{fig1}a), with that of a single QDM (fig \ref{fig1}b).  The form of the single dot data is well established, consisting of emission from a charge neutral exciton (\Xnot=1$e$+1$h$) that is gradually replaced by singly (\Xm=2$e$+1$h$) and doubly (\Xmm=3$e$+1$h$) negatively charged transitions as the electric field reduces and electrons transfer into the dot from the adjacent $n$-contact.  We note that the observation of \Xm~$\sim$3-5 meV below \Xnot~is very characteristic for self-assembled QDs \cite{warburton,baier,finley} reflecting the net attractive Coulomb interactions in the 3-particle (Trion) state.  
\\
In strong contrast, the QDM data consists of a rich spectrum of crossings and anticrossings, each of which occurs at a distinct electric field marked by vertical dashed lines on fig \ref{fig1}b.  This behavior arises from spin dependent quantum coupling of different initial (2$e$+1$h$, 1$e$+1$h$) and final (2$e$, 1$e$) states.  We begin with the neutral exciton anticrossing, labeled \Xnot~, at $F_{1e+1h}\sim$ 15.8 kV/cm.  As discussed in refs \cite{krennerNJP,krennerPRL} this arises from the quantum coupling of spatially direct ($e,h$ in upper dot) and indirect ($e$ in lower dot, $h$ in upper dot) states that are tuned into resonance due to the stronger quantum confined Stark shift of the latter.  Close to the anticrossing a peak labeled \textbf{indirect} \Xm~is observed between the two \Xnot~branches, following a characteristic \textit{s-like} trajectory with reducing field. The initial shift is parallel to the direct \Xnot~branch before stepping rapidly to lower energy between $\sim 7-11$ kV/cm and, thereafter, shifting parallel to \Xnot~again, precisely as predicted by Szafran \textit{et al.} for the behavior of the negatively charged Trion in a QDM\cite{szafran}. The s-like shift reflects a field driven redistribution of the \textit{electron} wavefunction amongst the upper and lower dots balanced by Coulomb interactions in the 3-particle initial state:  For singlet electron spin configuration \Xm~has \textit{direct} character at low field with both electrons and the hole localized in the upper dot (labeled \textbf{direct} \Xm~in fig \ref{fig1}b). This state evolves to have partially \textit{indirect} character as the field increases with one electron in the upper and lower dot.  The field range over which this \textit{direct} to \textit{indirect} transition occurs is governed by the interplay between the attractive and repulsive Coulomb interactions in the \Xm~state, the inter-dot quantum coupling energy and the relative energy of the orbital states in the upper and lower dots.\\
As expected, \Xm~ gains intensity as the electric field reduces whilst \Xnot~becomes weaker, indicative of electron charging in the limit of weak coupling to the contact\cite{baier}.  Furthermore, for $F\leq$ 7 kV/cm the \Xnot-\Xm~energy splitting tends to a constant value of  $\Delta E \sim 3.4$ meV, typical for the value in a single dot (see fig \ref{fig1}a and refs. \cite{warburton,finley}).  This is precisely the expected behavior for direct \Xnot~and \Xm~states in a QDM, where all particles in the initial state are localized within the upper dot.  We confirm this attribution below by calculating the spectrum of charged exciton transitions in the QDM.  Remarkably, these calculations also allow us to identify and explain \textit{all transitions} observed in fig \ref{fig1}b and obtain a full understanding of the behavior of $X^{-1}$ in a tunable QDM.

\begin{figure}
	\centering
		\includegraphics[width=0.90\columnwidth]{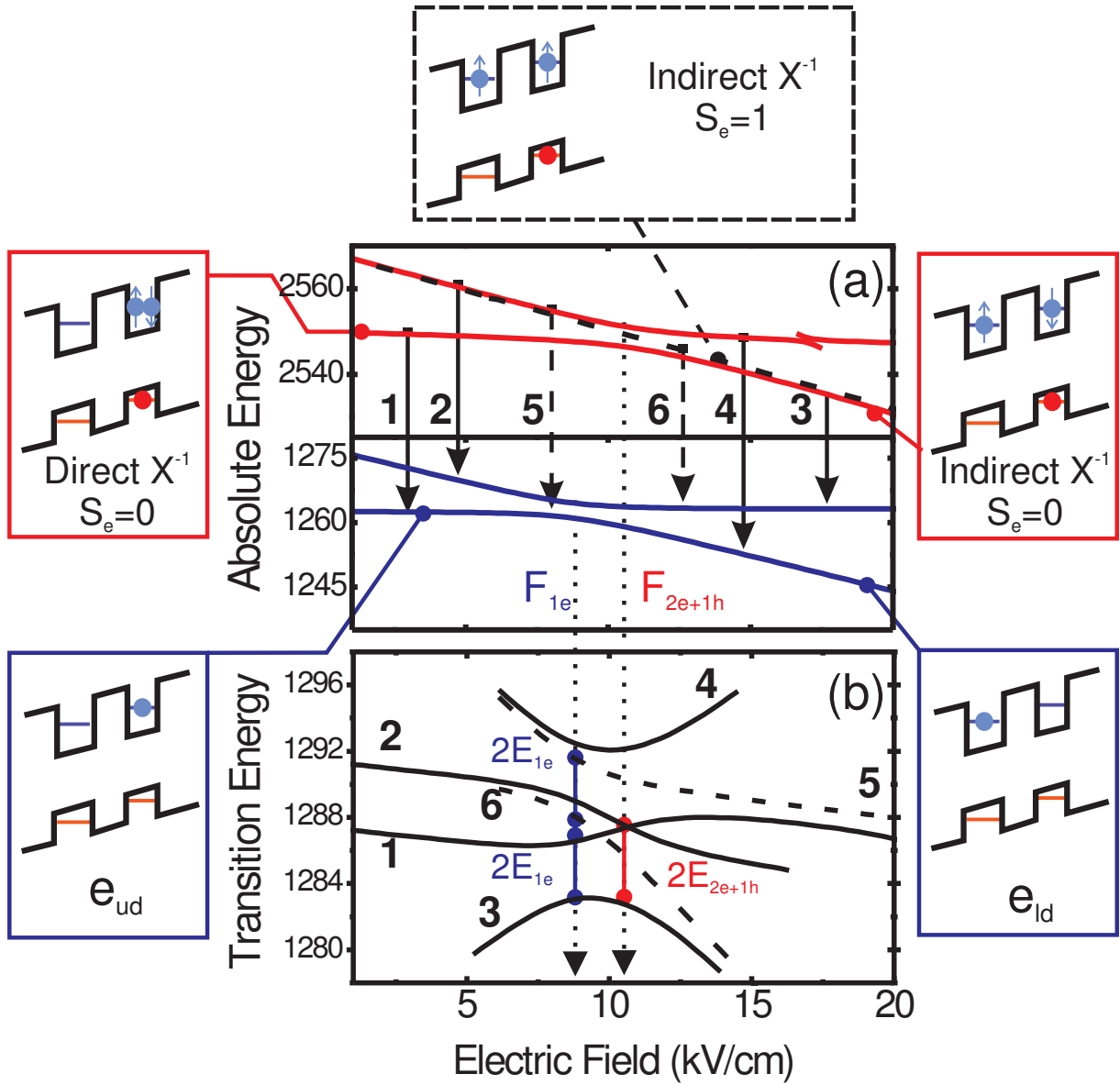}
	\caption{(color online) (a) Calculated absolute energies of initial (\Xm) and final $(1e)$ states as a function of electric field.  As illustrated by the schematics, singlet Trions with spatially direct and indirect character hybridize at $F_{2e+1h} = 10.6$ kV/cm whilst the triplet Trions do not hybridize in the initial state due to Pauli exclusion. Panel (b) shows the transition energies.}
	\label{fig2}
\end{figure}

We calculated the absolute energy of initial Trion ($2e+1h$) and the final $1e$ states as a function of electric field.  For these simulations we used a one band effective mass model\cite{burkard} that provides physical insight into the spin and charge dependent couplings without the need for detailed knowledge of the size, shape and composition of the QDM studied.  
Our results are presented in fig \ref{fig2}a with the associated transition energies plotted in fig \ref{fig2}b.  The schematics on fig \ref{fig2}a denote the spatial distribution of electrons and holes in \textit{unmixed} direct and indirect Trion initial and $1e$ final states.  As for \Xnot, direct and indirect Trions with singlet electron configuration (full lines) can be tuned into resonance by varying the electric field, anticrossing at $F_{2e+1h}$ = 10.6 kV/cm.  The energy splitting between the upper and lower Trion branches at resonance ($2E_{2e+1h}$ = 4.2 meV) reflects the tunnel coupling of two electrons in the \emph{presence} of the hole in the upper dot.  The final $1e$ state exhibits an anticrossing ($2E_{1e}$ = 3.2meV) at a different electric field ($F_{1e}$ = 8.6 kV/cm due) to the change of electrostatic energy of the QDM containing $1e$ as compared with $2e+1h$.  The dashed line on fig \ref{fig2}a shows the calculated energy of the spin triplet \Xm~state.  In strong contrast, only an \textit{indirect} triplet \Xm~state is energetically accessible over the investigated range of electric field, a manifestation of Pauli blockade phenomena.  Thus, the triplet Trion does not anticross in the initial state (dashed lines).  Optical transitions between the three initial Trion and two $1e$ final states are indicated by vertical arrows on fig \ref{fig2}a.\cite{stinaff} The energies of the resulting six transitions (labeled \textbf{1}-\textbf{6}) are presented in fig \ref{fig2}b, full and dashed lines corresponding to transitions from \emph{singlet} or \emph{triplet} Trion branches, respectively.\cite{parameters}  Transition \textbf{1} exhibits precisely the s-like trajectory discussed above in relation to fig \ref{fig1}b and can now be firmly identified as arising from the recombination of a singlet Trion with direct character for $F < F_{1e}$, mixed character for $F_{1e} < F < F_{2e+1h}$, and indirect character at higher field.  Transition \textbf{2} shows the opposite behavior corresponding to a singlet \Xm~state that has indirect character at low field but evolves into a direct state as the field increases (see fig \ref{fig2}a).  Transitions \textbf{3} and \textbf{4} both involve tunneling of the final state electron into a different dot during the recombination process.  As a result they are only expected to carry significant oscillator strength when either initial or final states have mixed character (i.e. $F_{1e} < F < F_{2e+1h}$).  This expectation is supported by the relative intensities of different transitions in fig \ref{fig1}b: Transition \textbf{3} is only observed for $F = 7-11$ kV/cm whilst transition \textbf{4} is not observed since it stems from the upper Trion branch (fig \ref{fig2}a) and phonon mediated population relaxation can occur over timescales comparable with the radiative lifetime.\cite{Nakaoka06}

\begin{figure}
	\centering
		\includegraphics[width=0.90\columnwidth]{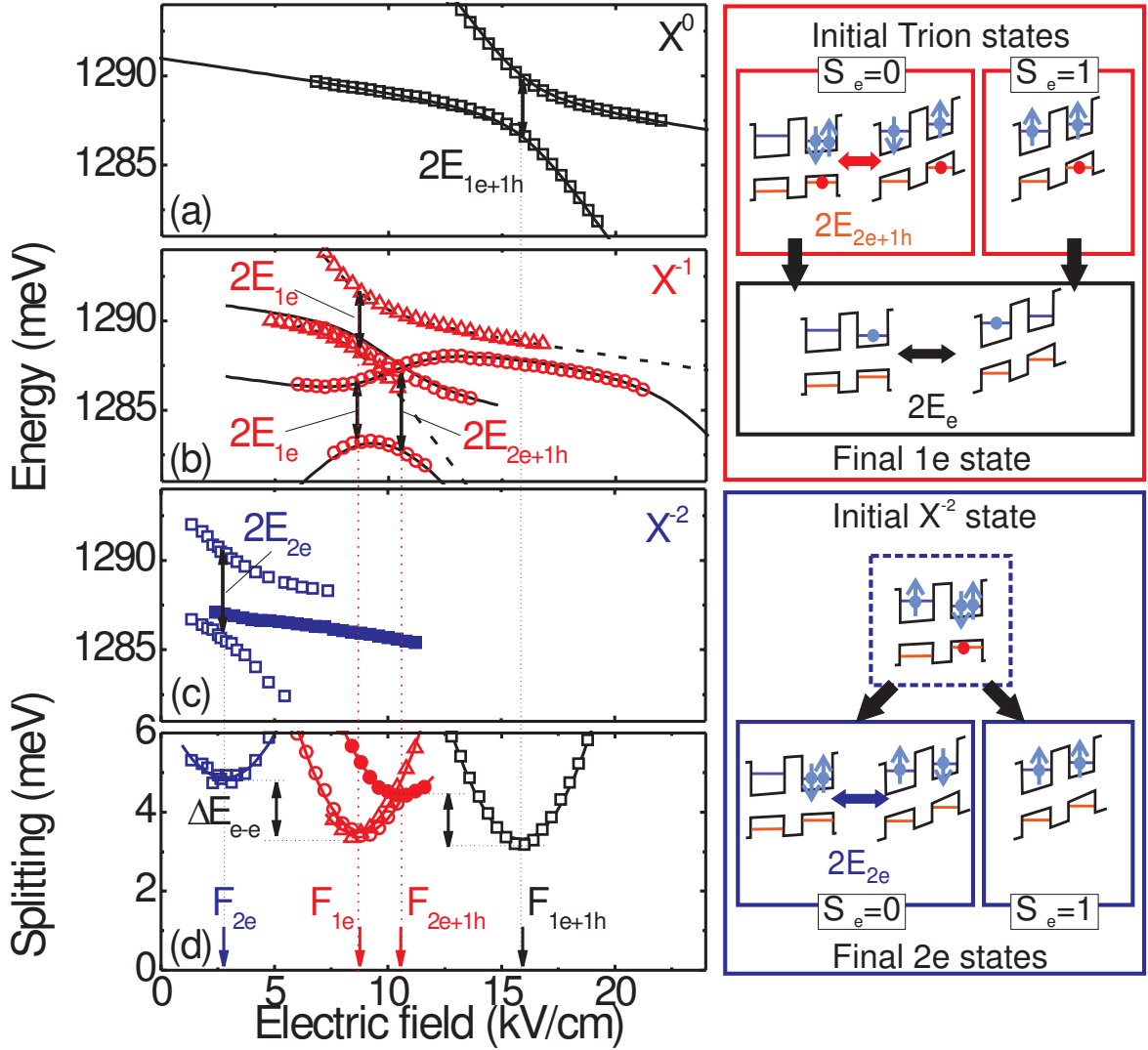}
	\caption{(color online) \textit{Left}: Comparison of experimentally measured peak energies (symbols) with calculated transition energies (lines) for \Xnot~(a), the singly \Xm~(b) and doubly \Xmm~(c) negatively charged excitons. (d) Measured energy splittings and fits of equation \eqref{eqn1} (solid lines).  \textit{Right}: Schematic representations of decay paths of \Xm~ (upper) and \Xmm~ (lower) transitions.}
	\label{fig3}
\end{figure}

Transitions from the \emph{triplet} \Xm~(\textbf{5},\textbf{6}) take place from a 3-fold degenerate initial state into two final 1$e$ states (fig \ref{fig2}). This is expected to give rise to an anticrossing at $F_{1e}$  that reflects the 1$e$ final state tunnel coupling (see fig \ref{fig2}b).  This expectation is unambiguously confirmed by our experimental data (fig \ref{fig1}b).  Figure 3 quantitatively compares our calculations with the measured peak positions for \Xnot~(fig \ref{fig3}a), singlet \Xm~(\textbf{1},\textbf{2},\textbf{3}) and triplet \Xm~transitions (\textbf{5},\textbf{6}) (fig \ref{fig3}b).  We find excellent agreement with experiment for all transitions using a single set of electron and hole confinement energies for the vertical and lateral motion in the upper and lower dots.\cite{parameters} 
\\
Further support for the peak identifications is obtained from the measured tunnel coupling energies for different few-particle states.  Transitions \textbf{1} and \textbf{3} take place from the same singlet \Xm~branch into different $1e$ final states.  As a result, the minimum energy splitting between \textbf{1} and \textbf{3} reflects the $1e$ tunnel coupling energy $2E_{1e}$.  Similarly, the minimum splitting between the triplet \Xm~peaks should also be equal to $2E_{1e}$ since the two transitions take place into the same pair of $1e$ final states.  The field dependence of the energy splitting $\Delta E$ between peaks \textbf{1}-\textbf{3} (open circles) and \textbf{5}-\textbf{6} (up triangles) are plotted in fig \ref{fig3}d and fit by:
 		\begin{equation}
 			\delta E = \sqrt{\left(2E\right)^2+\delta^2 \left(F-F_{crit}\right)^2},
 			\label{eqn1}
 		\end{equation}
where $2E$ is the tunnel coupling energy, $F_{crit}$ the critical field at the resonance and $\delta$ the electrostatic lever arm between the upper and lower dots.  Fitting equation \eqref{eqn1} to the observed splittings we obtain $2E = 2E_{1e} = 3.2\pm0.1$ meV, $F_{crit} = F_{1e} = 8.7\pm0.1$ kV/cm and $\delta  = 1.7\pm0.1$ meV/kV/cm for both singlet (\textbf{1}-\textbf{3}) and triplet (\textbf{5}-\textbf{6}) \Xm~splittings as expected.  This finding strongly supports the identification of peaks \textbf{1}-\textbf{3} and \textbf{5}-\textbf{6} as arising from recombination of negatively charged excitons having singlet and triplet spin character, respectively. In contrast, transitions \textbf{2} and \textbf{3} take place from different initial states into the same final $1e$ state.  Therefore, their splitting maps out the \Xm~initial state coupling energy $(2E_{2e+1h})$ and the anticrossing field for the Trion initial state. The energy splitting between peaks \textbf{2}-\textbf{3} is plotted in fig \ref{fig3}d (filled circles) together with a parabolic fit according to equation \eqref{eqn1}.
\\
The results of fitting \eqref{eqn1} to all initial (\Xnot~(1$e$+1$h$), \Xm~(2$e$+1$h$)) and final (1$e$) state anticrossings are summarized in table \ref{tab1}.  The coupling energies for $1e+1h$ initial and $1e$ final state anticrossings are found to be very similar ($2E_{1e+1h}$ = 3.4 meV and $2E_{1e}$ = 3.2 meV), confirming that the inter-dot tunnel coupling is mediated by the electron and the presence of the hole in the upper dot provides only a weak perturbation to the electron wavefunction.  Remarkably, adding a second electron to the QDM increases the coupling energy from 3.4 meV to 4.2 meV, i.e. by  $\Delta E_{e-e} = E_{2e+1h}-E_{1e+1h}$ = $1.2\pm0.1$ meV, since two particles now participate in the tunneling process.\cite{stinaff}
\\

\begin{table}
	\centering
	\begin{ruledtabular}
		\begin{tabular}{lccc}
			~                                        & $2E$                      &             $F_{crit}$            &   $\delta$  \\
			~                                        &  (meV)                               & (kV/cm)                  &   (meVcm/kV)\\\hline
			$1e+1h$ initial \Xnot                    &        $3.2\pm0.1$                   &   $15.85\pm0.1$                       &   $1.6\pm0.2$    \\ 
			$2e+1h$ initial \Xm                      &        $4.4\pm0.1$                   &   $10.6\pm0.1$                        &   $1.6\pm0.2$                 \\ 
			$1e$ final from \textbf{1}-\textbf{4}    &        $3.4\pm0.1$                   &   $8.5\pm0.1$                         &   $1.7\pm0.2$                 \\ 
			$1e$ final from \textbf{5}-\textbf{6}    &        $3.4\pm0.1$                   &   $8.9\pm0.1$                         &   $1.7\pm0.2$                 \\ 
		  			$2e$ final from \Xmm               &        $4.4\pm0.2$                   &   $2.8\pm0.1$                         &   $1.7\pm0.2$                 \\ 
		\end{tabular}
		\end{ruledtabular}		
		\caption{Measured coupling energies, resonant electric fields and lever arm from fitting equation \eqref{eqn1}.  For the presently investigated samples we expect $\delta= 1.5$ meVcm/kV, close to the measured value $\delta  = 1.6-1.7\pm 0.2$ meVcm/kV.}
		\label{tab1}
\end{table}

As shown on fig \ref{fig3}d, each few-particle configuration is tuned into resonance at a distinct value of the electric field.  This arises due to a change of the total electrostatic energy $(|\Delta E_{1e/1h}|)$ upon adding or removing charges and is a direct optical manifestation of inter-dot Coulomb blockade, independent of the tunnel coupling.  We calculated $|\Delta E_{1e/1h}|$ from the shift of the resonance field $(\Delta F_{1e/1h})$ as the QDM charge changes by $1e$ or $1h$. This is given by $|\Delta E_{1e/1h}|\sim(d+(h_{ud}+h_{ld})/2)\cdot|\Delta F_{1e/1h}|$, where $d$ = 10 nm is the inter-dot separation and $h_{ud(ld)}$ = 5 nm is the height of the upper (lower) dot. For example, moving between the $1e+1h$ (\Xnot) to $2e+1h$ (\Xm) anticrossings we measure $|\Delta F_{1e}| = F_{2e+1h}-F_{1e+1h} = 5.6\pm0.2$ kV/cm, corresponding to $|\Delta E_{1e}|$ = 8.7$\pm$0.3 meV.  Similarly, from the measured value of $|\Delta F_{1h}| = F_{1e}-F_{1e+1h}$ = 7.2$\pm$0.2 kV/cm we obtain $\Delta E_{1h}$ = 11.6$\pm$0.4 meV due to the addition or removal of $1h$.  Based on this analysis we expect the anticrossing of a 2$e$ state at $F_{2e}\sim$ 2.9 kV/cm, i.e. shifted by  $\Delta F_{1e}$ = -5.6 kV/cm relative to $F_{1e}$.  Remarkably, careful examination of the spectra in fig \ref{fig1}b does reveal an anticrossing at $\sim$ 2.8 kV/cm, in good agreement with this prediction.  This is generated by the decay of the lowest energy doubly charged exciton (\Xmm) with $2e+1h$ in the upper dot and 1$e$ in the lower dot.  As shown on fig \ref{fig3} \Xmm~can decay into $2e$ final states with singlet or triplet spin configuration.  As for the initial state of the singlet Trion, only the singlet couples in the final state giving rise to the observed 2$e$ anticrossing.  In contrast, if \Xmm~decays into the triplet 2$e$ final state, coupling is inhibited due to the large energy required to place one electron in the excited state of the upper dot.  This gives rise to a single emission line, precisely as observed in the spectra shown in fig \ref{fig1}b, midway between the two anticrossing branches (fig \ref{fig3}c).  By fitting equation \eqref{eqn1} to the two singlet \Xmm~lines we extract $2E_{2e}$ = 4.4$\pm$0.1 meV (fig \ref{fig3}d), equal to $2E_{2e+1h}$ (table \ref{tab1}) as expected for a double dot containing two electrons.
\\
In summary, we presented optical measurements of Coulomb- and Pauli-blockade effects in a single, electrically tunable, QD-molecule.  The tunnel couplings and electrostatic charging energies were independently measured for different few electron or negatively charged exciton states.  Inter-dot coupling was confirmed to be mediated by electron tunneling. Our findings were shown to be in very good agreement with realistic calculations that provide a complete description of the behavior of negatively charged excitons in quantum dot molecules.\\
\\
The authors gratefully acknowledge financial support by DFG via SFB 631 and one of us (TN) acknowledges financial support from the IT program (RR2002) from the MEXT (Japan).


\begin{thebibliography}\\
\bibitem {Petta} J. R. Petta, A. C. Johnson, J. M. Taylor, E. A. Laird, A.
  Yacoby \textit{et al}
  Science
  \textbf{309} 2180 (2005).
\bibitem{tarucha} S. Tarucha, D. G. Austing, T. Honda, R. G. van der Hage, and
  L. P. Kouvenhoven, Phys. Rev. Lett. \textbf{77}, 3613 (1996).
\bibitem{zrenner} A. Zrenner, E. Beham, S. Stufler, F. Findeis, M. Bichler, and G.
  Abstreiter, Nature (London) \textbf{418} 612 (2002).
\bibitem{stufler} S. Stufler, P. Ester, A. Zrenner, and M. Bichler, Phys. Rev.
  B \textbf{72} 121301(R) (2005).
\bibitem{krennerNJP} H. J. Krenner, S. Stufler, M. Sabathil, E. C. Clark, P. Ester,
  M. Bichler, G. Abstreiter, J. J. Finley and A. Zrenner, New J. Phys.
  \textbf{7} 184 (2005).
\bibitem{krennerPRL} H. J. Krenner, M. Sabathil, E. C. Clark, A.
  Kress, D. Schuh, M. Bichler, G. Abstreiter, and J. J. Finley, Phys. Rev.
  Lett. \textbf{94} 057402 (2005).
\bibitem{ortner} G. Ortner, M. Bayer, Y. Lyanda-Geller, T. L. Reinecke, A.
  Kress, J. P. Reithmaier, and A. Forchel Phys. Rev. Lett. \textbf{94} 157401
  (2005).
\bibitem{Kroutvar04} M. Kroutvar, Y. Ducommun, D. Heiss, M. Bichler, D. Schuh,
  G. Abstreiter and J. J. Finley, Nature (London) \textbf{432}
  81 (2004).
\bibitem{Elzerman04} J. M. Elzerman, R. Hanson, L. H. Willems van Beveren, B.
  Witkamp, L. M. K. Vandersypen and L. P. Kouwenhoven, Nature \textbf{430} 431
  (2004).
\bibitem{Troiani03} F. Troiani, E. Molinari, and U. Hohenester, Phys. Rev.
  Lett. \textbf{90} 206802 (2003).
\bibitem{Calarco03} T. Calarco, A. Datta, P. Fedichev, E. Pazy, and P. Zoller,
  Phys. Rev. A \textbf{68} 12310 (2003).
\bibitem{szafran} B. Szafran, T. Chwiej, F. M. Peeters, S. Bednarek, J.
  Adamowski, and B. Partoens, Phys. Rev. B \textbf{71}, 205316 (2005).
\bibitem{warburton} R. J. Warburton, C. Schäflein, D. Haft, F. Bickel, A.
  Lorke, K. Karrai, J. M. Garcia, W. Schoenfeld and P. M. Petroff, Nature
  (London) \textbf{405} 926 (2000).
\bibitem{stinaff} E. A. Stinaff, M. Scheibner, A. S. Bracker, I. V. Ponomarev,
  V. L. Korenev, M. E. Ware, M. F. Doty, T. L. Reinecke, and D. Gammon
  , Science, \textbf{311} 636 (2006).
\bibitem{baier} M. Baier, F. Findeis, A. Zrenner, M. Bichler, and G.
  Abstreiter, Phys. Rev. B \textbf{64} 195326 (2001).
\bibitem{finley} J. J. Finley, P. W. Fry, A. D. Ashmore, A. Lemaitre, A. I.
  Tartakovskii \textit{et al}
  
  Phys. Rev. B \textbf{63} 161305 (2001).
\bibitem{burkard} G. Burkard, G. Seelig, and D. Loss, Phys. Rev. B \textbf{62}
  2581 (2000).
 
\bibitem{parameters} The best fit to our experimental data was obtained with:
		$m^*_{e,(x,y)/z}=0.07$,$m^*_{h,z}=0.35$,$m^*_{h,(x,y)}=0.19$ in units of $m_0$.  Interdot separation $d=9$ nm.  The characteristic energies of the parabolic confinement potentials used were 
		$\hbar\omega_{e,(x,y)}=$23meV, 
    $\hbar\omega_{e,(z)}=$48meV, 
    $\hbar\omega_{h,(x,y)}=$16meV,
    $\hbar\omega_{h,(z)}=$34meV,
   
\bibitem{Nakaoka06} T. Nakaoka, H. J. Krenner, E.C. Clark, M. Sabathil, M. Bichler, Y. Arakawa, G. Abstreiter and J. J. Finley.  Submitted for Phys. Rev. B, Preprint at \texttt{cond-mat/0607023}

\end{thebibliography}
\end{document}